\documentclass[
twocolumn,
]{ceurart}

\begin{document}

\copyrightyear{2022}
\copyrightclause{Copyright for this paper by its authors.
  Use permitted under Creative Commons License Attribution 4.0
  International (CC BY 4.0).}

\conference{CIKM 22: Workshop on Human-In-the-Loop Data Curation, October 21, 2022, Atlanta, GA}

\title{Knowledge Management System with NLP-Assisted Annotations: A Brief Survey and Outlook}


\author[1]{Baihan Lin}[%
orcid=0000-0002-7979-5509,
email=baihan.lin@columbia.edu,
url=https://www.neuroinference.com/,
]
\cormark[1]
\address[1]{Columbia University,
  New York, NY 10027, USA}

\cortext[1]{Corresponding author.}

\begin{abstract}
Knowledge management systems (KMS) are in high demand for industrial researchers, chemical or research enterprises, or evidence-based decision making. However, existing systems have limitations in categorizing and organizing paper insights or relationships. Traditional databases are usually disjoint with logging systems, which limit its utility in generating concise, collated overviews. In this work, we briefly survey existing approaches of this problem space and propose a unified framework that utilizes relational databases to log hierarchical information to facilitate the research and writing process, or generate useful knowledge from references or insights from connected concepts. Our framework of bidirectional knowledge management system (BKMS) enables novel functionalities encompassing improved hierarchical note-taking, AI-assisted brainstorming, and multi-directional relationships. Potential applications include managing inventories and changes for manufacture or research enterprises, or generating analytic reports with evidence-based decision making. 
\end{abstract}


\maketitle

\section{Introduction}

Knowledge management systems (KMS) are the driving engines of modern day information technologies (IT). These IT systems store data in parsed ways and retrieve knowledge insights to improve the information understanding, team collaboration and process alignment within organizations and groups. As an engineering entities in high demand for industrial researchers, chemical or research enterprises and evidence-based decision making, knowledge management systems are often used by organizations to affect innovation performance and generate accurate metrics on organizational capacity \cite{lawson2001developing}, but they can also be user-centric by centering the knowledge base around individual users or customers \cite{kabir2014user}. 

Take the application of reference management of academic researchers as an example. KMS are often used by researchers to keep track of papers or subsets of papers \cite{yee2019back}. Usually, the research information of different papers or references has meta information that can be filtered and sorted. An example scenario would be: a scientist logs or inputs a particular paper into a system, with each entry containing many meta information about the papers. These meta information elements can be filtered or sorted (e.g., by year, journal, author, etc.). Each paper might contain multiple concepts or topics, and each topic might contain multiple paper. In some cases, we might want the system to be able to automatically assign topic to some papers based on text data mining. The user can filter the papers by topics. Within each paper, during the reading, the scientist might want to log an insight or note on certain paragraphs. Sometimes the notes can be about multiple papers, and their relationship can be in various types. These notes or insights also have topic tags, which can optionally be automatically curated. The system can also generate useful concepts or knowledge as well as their references to facilitate the research and writing process of the scientist.

We see from this example that the relationships between papers chosen in academic fields can have multiple, bidirectional relationships. Existing knowledge management systems for organizing research papers in scientific fields or organizing manufacture enterprises use directed acyclic graphs, Bayesian networks, and machine learning \cite{yee2019back}, which have limitations in categorizing and organizing these multi-faceted insights or relationships. This is because many traditional databases are usually disjoint with logging systems, which limit its utility in generating concise, collated overviews. In this work, we briefly survey existing approaches in the general field of these knowledge management systems, and propose a unified framework as a solution to these challenges. In our framework, we describe a knowledge management system that utilizes relational databases to log hierarchical information with connected concepts. 

Back to the example problem of reference management, our KMS would utilize relational databases to log hierarchical information to facilitate the research and writing process, or to help generate useful knowledge from references or insights from connected concepts. This would enable novel functionalities encompassing improved hierarchical notetaking, AI-assisted brainstorming, and multi-directional relationships. For instance, one can generate reports given keywords or topics collating hierarchical and intra-connected records. With these automatic annotations, the system can enable automatic curation of topic tags using text data mining. Other applications include managing inventories and changes for manufacture or research enterprises or generating analytic reports with evidence-based decision making. 

Although we have seen successful system designs in commercial products such as Mendeley and recent community efforts such as Open Research Knowledge Graph (ORKG), we believe that our survey can still bring useful and new insights on the practical considerations on the intersections among machine learning, database management and human-system collaboration. In the following sections, we will first briefly survey the existing knowledge management systems approaches, and propose a unified bidirectional KMS (BKMS) framework that utilizes relational databases to log hierarchical information to facilitate the research and writing and generate helpful knowledge from references or insights from related concepts. We present a useful and novel system design for this bidirectional information management, formulate a few potential use-cases for this design, address the four-subset system of NLP-assisted annotations, and discuss future design considerations.

\section{An Applied Perspective}

\subsection{Applications}

There are different application domains for knowledge management systems with relational databases and insight annotation enabled by machine learning, including but not limited to reference manager for academic researchers, education and research tool, consulting firm report generator with evidence-based decision making, inventory management for manufacture or research enterprises, organizational tool for industries with high-volume data, and internal auditing tool for customized employee metrics. 

\subsection{User scenarios}

Other than the reference management example in our introduction, we also include two additional applications. The first one is managing inventories and changes for manufacture, chemistry or research enterprises. The inventories or measurements of factories usually involves dependency and hierarchical interactions. A knowledge management system that uses a relational database instead of disjoint databases with separate logging systems can enable useful curation function to offer very useful and concise report regarding key events or phenomon (like the topics). These are important insights to keep the factories or warehouses in safety.

The second user scenario example is evidence-based decision making. In large business entities, critical decisions are usually made with a group of market researchers or consulting firms that come up with various analytic reports. A knowledge management system with AI-assisted insight annotation can provide a fast and evidence-based solution by generating a report (given the keyword or topic as input) which curates from hierarchical and interaconnected records. This hierarchical knowledge graph can serve as a useful primer in important decision making processes and guide the investigators to locate relevant resources.

\subsection{Case studies}

In this section, we outline three case studies that recent real-world knowledge management systems are likely adopt to become more interconnected and intelligent. 

\textit{The concept of Internet of Things (IoT)}: The IoT advancements consist of a series of disruptive digital technologies, semantic languages, and virtual identities that can increases efficiency and effectiveness in daily life operations through interconnected communications among devices and systems \cite{scuotto2016internet}. Other than these organizational benefit, IoT stimulates the innovation process in various aspects, through fast iterations of knowledge flow and information gathering \cite{malhotra2000knowledge}. In \cite{santoro2018internet}, researchers employ structural equation modelling on a sample of 298 Italian firms from different sectors. Their study suggest that interconnected knowledge management systems facilitate the creation of a open and collaborative ecosystem by utilizing the internal and external flows of knowledge and increasing internal knowledge management capacity, which in turn increases innovation capacity. 


\textit{Reference architecture}: In the era of Industry 4.0 \cite{lasi2014industry}, smart warehouses are envisioned to host production that contains modular and efficient manufacturing systems and characterizes scenarios in which products control their own manufacturing process. As in our user scenario of warehouse inventory management, an optimal reference architecture would be the key to the warehouse knowledge management system. For instance, \cite{van2021design} describes a pipeline to perform a series of systematic analyses to identify the key concerns and processes and eventually arrive at potential architecture of smart warehouses. They conduct a case study at a large warehouse in the food industry and illustrates that an introduction of a reference architecture can be effective and practical. 

\textit{Conversational recommendation systems}: 
A conversational recommendation system (CRS) is a computer system that is able to have a conversation with a human user in order to make recommendations \cite{sun2018conversational}. This is different from traditional recommendation systems, which do not interact with users. Often used in e-commerce, social media, and entertainment applications, CRS are becoming increasingly popular as they can provide a more personalized and interactive experience for users, but can pose additional challenges in managing different layers of knowledge at different states: the intent of the conversation, the entities matched by the intents, the long-term preferences of the users and similar users, their state-dependent preferences related to the current contexts, and the relationships between different entities, intents and users. One practical examples is recommending discussion topic to therapist during psychotherapy in real-time given automatically speech-transcribed dialogue records \cite{lin2022supervisor} and helpful visual analytics \cite{lin2022voice2alliance}.

\begin{figure*}[tb]
  \includegraphics[width=\linewidth]{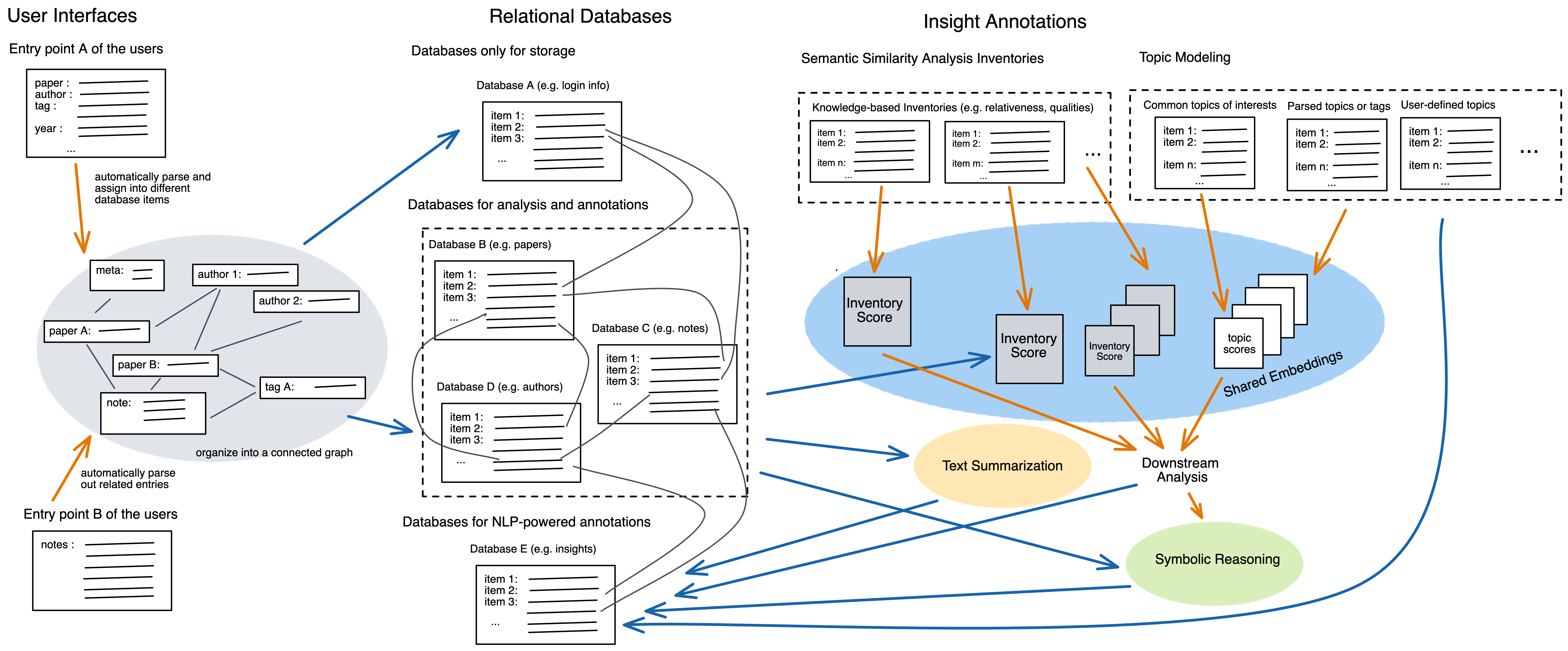}
  \caption{A unified framework of a knowledge management system with relational databases and NLP-assisted annotation}
  \label{fig:teaser}
\end{figure*}

\section{Bidirectional KMS Framework}

Figure \ref{fig:teaser} outlines our framework of bidirectional knowledge management systems (BKMS) with relational databases and insight annotation powered by natural language processing (NLP). The user interface provides the entry points into our knowledge management systems. Different interfaces introduces different routes, but they all involve a parsing and extraction process to atomize the user inputs into nodes that connects in a small knowledge graph. This graph is then placed into a relational database where their links are preserved. The orange and blue arrows indicates intro- and inter-database data flows. The relational databases include three parts. Some databases in the relational databases are only used for storage. Some are used for analysis and annotations. And some databases are kept to store annotated insights or other downstream analytical artifacts, which provide an additional data flow direction. 

\section{NLP-Assisted Insight Annotation}

As shown in the annotation component of Figure \ref{fig:teaser}, there are several routes we can utilize natural language processing to generate and annotate insights within our databases. We will elaborate on how they play in knowledge management systems and survey modern machine learning methods in each of these routes below.

\textit{Semantic similarity:}
In principle, any sentence or paragraph embeddings can help us characterize our document and inventories of interest. For instance, the Doc2Vec embedding \cite{le2014distributed} is a popular unsupervised learning model that learns vector representations of sentences and text documents. It improves upon the traditional bag-of-words representation by utilizing a distributed memory that remembers what is missing from the current context. SentenceBERT \cite{reimers2019sentence} is another popular option which modifies a pre-trained BERT network by using siamese and triplet network structures to infer semantically meaningful sentence embeddings. With word or sentence embeddings, we can embed the document entries from our relational databases into vectors, and then compute the cosine similarity between the vector at certain turn and an inventory entry. With that, for each text, we obtain a \textit{N}-dimension score for the said property. For instance, the inventory can be written guidelines that evaluate the usefulness of certain documents, say, a list of leadership principles that some companies use to evaluate a candidate's resume, work report or performance review form. And the relational database could be hosting an employee's self reported performance review form. The system can automatically compute a score based on each item of the guidelines and annotate these document entry accordingly. Other applications can be evaluating the patient-doctor alignment from an automatically transcribed psychotherapy sessions based on a clinical questionnaire inventory, as shown in \cite{lin2022deep,lin2022mbti,lin2022deep2}.

\textit{Topic modeling:}
In natural language processing and machine learning, a topic model is a type of statistical graphical model that help uncover the abstract ``topics'' that appear in a collection of documents. The topic modeling technique is frequently used in text-mining pipeline to unravel the hidden semantic structures of a text body. This can be very handy in annotating the database entry. For instance, a user scenario could be in a clinical consumer-facing chatbot, where the dialogue between the client and agent is transcribed, and a topic modeling analysis is automatically performed and generate a list of discussed topics and their scores based on semantic similarity, as shown in \cite{lin2022neural}. Several state-of-the-art neural topic models include the Neural Variational Document Model (NVDM) \cite{miao2016neural} (an unsupervised text modeling approach based on variational auto-encoder), Gaussian softmax construction (GSM) \cite{miao2017discovering} (a NVDM variant), the Wasserstein-based Topic Model (WTM) \cite{nan2019topic}, the Embedded Topic Model (ETM) \cite{dieng2020topic} among others. 

\textit{Text summarization:}
When the scale of our databases increases, maintaining the interpretability of our knowledge management system becomes more and more challenging. This expanding availability of documents and entries inside the database cannot yield actionable insights without proper aggregation. The field of automatic text summarization deals with this problem by producing a concise and fluent summary while preserving key information content and overall meaning \cite{allahyari2017text}. For instance, we can first group or cluster the database entries (such as paper abstracts, or reading notes as in our reference manager example) by their semantic similarity or inferred topics. And then, within each group, generate a condensed descriptions. A user case would be, automatically generating writing outlines or topics based on the available references and reading notes in a paper reference manager. In the active field of text summarization, extraction and abstraction are the two main approaches. The extractive summarization techniques generate summaries by choosing a subset of the sentences in the original text, by computing first an intermediate representation of the text, then a sentence score and finally a subset selection operation onto the original texts \cite{nenkova2012survey}. The abstraction approach uses latent semantic analysis, frequency-driven approaches \cite{dunning1993accurate} and topics modeling which we cover above. 

\textit{Symbolic reasoning:}
While topic modeling offers interpretable subjects, and text summarization offers interpretable paragraphs, the logic and causal relationship between these insights can be arbitrary. The field of symbolic AI bridge this gap by introducing high-level and human-readable symbolic representations into these practical problems. They can potentially derive logic programming rules and semantic relationships that can be use as actionable knowledge graphs \cite{garnelo2019reconciling}.
Recently, there have also been increasing interests in a modern approach called neuro-symbolic AI \cite{garcez2020neurosymbolic,zhang2021neural}, where the well-founded knowledge representation and reasoning from the symbolic perspective are integrated with deep learning from the statistical perspective. This offers both effective predictive power and necessary explainability for many real-world applications.


\section{Practical Considerations}

When designing a interconnected and intelligent knowledge management systems for a domain-specific application, here are some practical questions to be considered: 

\begin{itemize}
    \item \text{Database consideration}: What are the storage capacities of this technology?
    \item \textit{User interface}: What visual and user interface is preferred by users?
    \item \text{Organizational benefits}: What specific organizational functionality would this system provide over current systems?
    \item \text{Latency and responsiveness}: What are the synchronization capacities of this technology across devices?
    \item \text{Customization}: Can users modify or customize this system to their own preferences?
    \item \text{Security}: Would this technology allow for secure encryption or storage of higher value data?
    \item \text{Collaboration}: Would this system allow for collaborative use by multiple stakeholders?
    \item \text{Investigation}: What kind of insights or investigations do we wish to gain from this system?
    \item \text{I/O}: Would this system allow import or export from other knowledge management systems?
\end{itemize}

Other than these practical questions to consider, a more thorough design process would involve market analysis (market size, emerging technologies, policies, challenges, new trends, and policies as in \cite{giudici2020cryptocurrencies}), domain analysis (systematic activity for deriving, storing domain knowledge to support the engineering design process as in \cite{koksal2017feature}), business process modeling (i.e. identifying the lead processes and subprocess of outgoing products \cite{weske2019business}) and architecture design with viewpoints (stakeholder concerns, context diagram, decomposition view, uses view, and deployment view \cite{clements2003documenting,demirli2011software}). Sometimes, case studies can also be useful to clarify the problem settings. 

Since we are proposing the idea of introducing relational databases and various AI and symbolic techniques in knowledge management systems, there are additional future research challenges in relation to this proposition in terms of the human-system ``collaboration'' enabled by these systems. Methodologically, tne machine learning engine that powers many human-in-the-loop (HIL) solutions in data curation is reinforcement learning methods that have been demonstrated to effectively learn from human interactions with the speech- or text-based systems \cite{lin2022reinforcement}. Operationally, from the human side, we need to encourage people to contribute their knowledge and expertise (e.g. crowdsourcing) by creating an effective user interface that allows people to easily log in, search for and find the information they need.From the system side, we need to ensure that knowledge is effectively captured and stored, consistently updated to keep the knowledge up to date and accuratem and manage different types of knowledge such that it is accessible to the right people. Finally, there are also ethical and societal considerations when we use machine learning and AI to encode knowledge related to human biometrics and well-beings, as reviewed in \cite{lin2022computational}.

\section{Conclusions}

In summary, we describe the applied problem of a knowledge management systems that host information that contain multiple and bidirectional relationships in layers of meta data. We briefly survey the application domains, user scenarios and the existing approaches in the fields, and eventually propose a framework for a knowledge management system with relational database and NLP-assisted insight annotation.
In our framework, a knowledge management system can comprise a user interface to provide input and present output relating to one or more documents or sensors. The system maintains a relational database storing information relating to the one or more documents, and a knowledge parsing unit, in communication to the user interface and the server, can determine at a first time instance the metadata information elements associated with the particular document entry. The databases can then be automatically annotated with NLP techniques such as semantic similarity analysis, topic modeling, text summarization and symbolic reasoning. A knowledge graph can then be learned from these language models to be used as interpretable insights for real-world downstream tasks.


\bibliography{main}




\end{document}